\documentclass[a4paper]{amsart}
\usepackage{a4wide}
\usepackage{amsthm, amsmath, amssymb, amsopn}
\usepackage{hyperref}
\usepackage{mathrsfs,comment}
\usepackage{xcolor}
\usepackage{thm-restate}
\usepackage[normalem]{ulem}
\declaretheorem[numberwithin=section]{theorem}

\newtheorem{proposition}[theorem]{Proposition}
\newtheorem{lemma}[theorem]{Lemma}
\theoremstyle{definition}
\newtheorem{definition}{Definition}
\theoremstyle{remark}
\newtheorem{example}{Example}

\DeclareMathOperator\csp{CSP}
\DeclareMathOperator\typ{typ}

\DeclareMathOperator\Clo{Clo}
\DeclareMathOperator\Hom{Hom}

\newcommand\ignore[1]{}
\newcommand{\sig}[1]{\ensuremath{#1}}           %
\newcommand{\alg}[1]{\ensuremath{\mathbf{#1}}}  %
\newcommand{\rel}[1]{\ensuremath{\mathscr{#1}}} %
\newcommand{\mbf}[1]{\ensuremath{\mathbf{#1}}}

\DeclareMathOperator{\graph}{Gr}

\newcommand\homcsp[1]{\ensuremath{\csp(#1)}}

\newcommand\counting[1]{\ensuremath{c_{#1}}}
\newcommand\Kpoly{\ensuremath{\mathcal K_{poly}}}
\newcommand\Kpolyeff{\ensuremath{\mathcal K_{eff}}}
\newcommand\Ksurj{\ensuremath{\mathcal K^{s}_{poly}}}
\newcommand\Ksurjeff{\ensuremath{\mathcal K^{s}_{eff}}}

\author{Libor Barto\and William DeMeo\and Antoine Mottet}
\address{Department of Algebra, Faculty of Mathematics and Physics, Charles University in Prague}
\email{libor.barto@gmail.com, williamdemeo@gmail.com, mottet@karlin.mff.cuni.cz}

\begin{document}

\title{Constraint Satisfaction Problems over Finite Structures}

\maketitle

\begin{abstract}
We initiate a systematic study of the computational complexity of the Constraint Satisfaction Problem (CSP) over finite structures that may contain both relations and operations. We show the close connection between this problem and a natural algebraic question: which finite algebras admit only polynomially many homomorphisms into them?

We  give some sufficient and some necessary conditions for a finite algebra to have this property.
In particular, we show that every finite equationally nontrivial algebra has this property which gives us, as a simple consequence, a complete complexity classification of CSPs over two-element structures, thus extending the classification for two-element relational structures by Schaefer (STOC'78).

We also present examples of two-element structures that have bounded width but do not have relational width (2,3), thus demonstrating that, from a descriptive complexity perspective, allowing operations leads to a richer theory.
\end{abstract}

\section{Introduction}\label{introduction}

Constraint satisfaction problems (CSPs) and their variants form a natural class that includes a vast array of practical problems studied in computer science.
For the traditional decision variant of the so-called \emph{non-uniform CSP}, each problem is parametrised by a relational structure $\rel A$ with a finite signature, called the \emph{template}, and the problem is to decide whether a given input $\rel X$ admits a homomorphism to $\rel A$. We denote this problem by $\homcsp{\rel A}$. It is known that every decision problem is polynomial-time Turing equivalent to $\homcsp{\rel A}$, for some suitably chosen infinite structure $\rel A$~\cite{MR2503587}.
For finite structures, the problem is considerably better behaved; in particular, the CSP over every finite structure is clearly in NP. On the other hand, finite-template CSPs still form a rather broad class that includes variants of Boolean satsfiability problems, graph coloring problems, and solving equations over finite algebraic structures.

A classic result in the field is a theorem of Schaefer~\cite{Schaefer:1978} that completely classifies the complexity of CSPs over relational structures with a two-element domain (so called \emph{Boolean structures}) by providing a dichotomy theorem: each such CSP is either in P or NP-complete. This result was one of the motivations for a famous dichotomy conjecture of Feder and Vardi~\cite{Feder:1999} stating that the dichotomy extends to arbitrary finite domains. 
Their conjecture initiated the rapid development of a general theory of non-uniform finite-domain CSPs, called the \emph{algebraic approach} to CSP, culminating in a positive resolution by Bulatov~\cite{Bulatov:2017} and, independently, Zhuk~\cite{Zhuk:2017, Zhuk:2020}.

The general theory was extended with many variants and generalizations of finite-template CSPs, and this has led to many new results. For example, the complexity was classified within large classes of infinite-domain CSPs~\cite{Mottet-MMSNP,Mottet-temporal,Pinsker-graphs}, optimization problems~\cite{KKR17}, and promise problems~\cite{Brakensiek:2018,Ficak:2019}.

In this paper we study a generalization of finite-template CSP that goes in a different direction than those above: we study the complexity of CSPs over arbitrary finite structures, that is, we allow relation symbols \emph{and operation symbols} in the signature (as opposed to classical CSPs which admit relation symbols only). We take the first steps toward an algebraic approach to these problems,
providing in particular a generalization of Schaefer's dichotomy theorem from Boolean relational structures to Boolean (general) structures, and demonstrating that this class of problems is richer in the sense of descriptive complexity. 

Before we present the contributions in detail, let us first describe our motivations for this line of research. 

\subsection{Motivation}
One of our motivations is straightforward. Allowing function symbols in the signature is a very natural generalization, particularly since the modern theory of purely relational CSPs has a     manifestly algebraic character. On the other hand, a general theory for CSPs over arbitrary structures is missing. The only work that studies such problems seems to be the paper by Feder, Madelaine, and Stewart~\cite{Feder:2004} which shows, in particular, that even the purely algebraic setting generalizes the relational one in the sense that each CSP over a finite relational structure is equivalent to a CSP over a finite algebraic structure modulo polynomial-time Turing reductions.

A deeper motivation for studying CSPs over arbitrary structures is that they can be regarded as a sort of hybrid restriction of CSPs over relational structures, as we explain in the next section.

Non-uniform CSPs can be viewed as one extreme of the general, or uniform, homomorphism decision problem.  The input of such a problem is a pair $(\rel X, \rel A)$ of structures and one must decide whether there exists a homomorphism from $\rel X$ to $\rel A$.  Non-uniform CSPs, on the other hand, are right-hand side restrictions (also known as \emph{language} restrictions) in that the target (or codomain) $\rel A$ is fixed or restricted to belong to some fixed class of structures. At the other extreme, left-hand side or \emph{structural} restrictions, where the source (or domain) structure is fixed, or restricted to a special class of structures,%
\footnote{Note here that fixing $\rel X$ to a single structure gives a problem that can always be solved in polynomial time (in the size of \rel A).} also have a highly-developed theory~\cite{Marx:2013} which seems, however, quite different from non-uniform CSPs at present.

The study of \emph{hybrid restrictions}, where both left- and right-hand sides are restricted may help in bridging the gap between the two extreme cases.
Such hybrid restrictions have been studied in the literature.
Dvo\v{r}\'ak and Kupec~\cite{Dvorak:2015} studied the so-called \emph{planar Boolean CSPs},  where $\rel A$ is a fixed 2-element structure and where the instances $\rel X$ are required to be planar.
Kazda, Kolmogorov, and Rol\'inek~\cite{Kazda:2019} studied the case where $\rel A$ is a fixed 2-element structure and where there is a bound $k$ such that every element of $\rel X$ appears in at most $k$ tuples in a relation of $\rel X$.
This work settled a question posed by Dvo\v{r}\'ak and Kupec and established that there is a P/NP-complete dichotomy for planar Boolean CSPs.
However, a general theory still seems far off.

A natural hybrid restriction on CSPs of relational structures is to require that some of the constraints in the instances are graphs of functions.
This type of hybrid restriction is precisely what the framework of CSPs of arbitrary structures captures.
Indeed, every structure $\rel A$ can be turned into a relational structure by considering its graph, denoted by $\graph(\rel A)$.
In this way, the restricted instances of $\csp(\graph(\rel A))$ correspond precisely to the instances of $\csp(\rel A)$.
As we show below, the study of the complexity of CSPs of arbitrary structures is amenable to algebraic methods, giving natural complexity invariants for this particular type of hybrid restriction.

\subsection{Complexity Invariants}

The seminal discovery in the algebraic approach to CSP is the result of Jeavons~\cite{Jeavons:98} that, for a finite relational structure \rel A, the complexity of $\csp(\rel A)$ is completely determined by the set of \emph{polymorphisms} of $\rel A$---the functions from $A^n$ to $A$ that preserve all the relations of $\rel A$. This is because the complexity of $\csp(\rel A)$ does not increase when $\rel A$ is expanded by primitively positively (pp-) definable relations, and this is exactly what polymorphisms control.

For general finite structures we show in Section~\ref{sect:general} that $\rel A$ can be expanded,  without increasing the complexity, by operations defined by  terms from the operations in $\rel A$ and \emph{quantifier-free pp-definable} relations (Definition~\ref{def:qfpp}). This yields the following complexity invariant.

\begin{restatable}{theorem}{invariant} \label{thm:general}
	For a finite structure $\rel A$, the complexity of $\csp(\rel A)$ depends only on the clone generated by the basic operations of the algebraic reduct of $\rel A$
	together with the set of partial polymorphisms of the relational reduct of $\rel A$.
\end{restatable}

We do not know whether quantified pp-definable relations can be safely added as well when operations are present; in other words, we do not know whether ``partial polymorphisms'' can be replaced by ``polymorphisms'' in this theorem.

The discovery of the crucial role of polymorphisms for relational CSPs has been followed by improvements~\cite{Krokhin:2005, Barto:2018} which made it possible to conjecture the borderline between P and NP-complete CSPs over non-Boolean domains and which were important for further generalizations to, e.g., infinite domains \cite{Bodirsky:2006}, promise CSPs \cite{AGH17,Brakensiek:2016,Bulin:2019}, and valued CSPs~\cite{Cohen:2013}.
In the non-relational setting, we do not yet have analogues to all of these advances in relational CSPs. However, Lemma~\ref{lem:eq} is a step in this direction and plays an important role in the results we present here.

\subsection{Boolean structures}

The complexity invariants were obtained on the way to one of our goals, which was to generalize Schaefer's dichotomy theorem from relational structure to arbitrary structures with a two-element domain.

In~\cite{Feder:2004}, Feder, Madelaine, and Stewart prove that for a structure $\rel A$ whose operations are at most unary, the CSP over $\rel A$ is equivalent to the CSP over $\graph(\rel A)$ under polynomial-time many-one reductions. An easy observation is that this fact extends to operations of essential arity at most 1 (by Lemma~\ref{lem:general-reduction}, for example) and therefore the complexity classification follows from Schaefer's result in this case.
We show that in all other cases (that is, when an operation of essential arity at least 2 is present) the CSP is solvable in polynomial time (Theorem~\ref{thm:boolean-dichotomy}).

\begin{restatable}{theorem}{booleandichotomythm} \label{thm:boolean-dichotomy}
Let $\rel A$ be a Boolean structure of finite signature. 
\begin{itemize}
    \item If all functions in $\rel A$ are essentially unary, then
      $\csp(\rel A)$ is equivalent to $\csp(\graph(\rel A))$.
    \item If $\rel A$ contains an operation that is not essentially unary, 
      then $\csp(\rel A)$ is in P.
\end{itemize}
\end{restatable}

Schaefer's dichotomy can be phrased using polymorphisms as follows: if $\rel A$ is a relational structure, then $\csp(\rel A)$ is in P if $\rel A$ has a constant polymorphism or has a polymorphism that is not essentially unary, and it is NP-complete otherwise.
Thus Theorem~\ref{thm:boolean-dichotomy} can be phrased as follows.

\begin{restatable}{corollary}{booleandichotomy}
\label{cor:boolean-dichotomy}
Let $\rel A$ be a Boolean structure of finite signature. 
If the algebraic reduct of $\rel A$ contains an operation that is not essentially unary or
the graph of $\rel A$ admits a constant polymorphism or a polymorphism that is not essentially unary, then $\csp(\rel A)$ is in P. Otherwise, $\csp(\rel A)$ is NP-complete.
\end{restatable}

\subsection{Polynomially-Many Homomorphisms}

It turned out, and it was surprising to us, that \emph{all} the CSPs in the second item of Theorem~\ref{thm:boolean-dichotomy} can be solved in polynomial time for a particularly strong reason,
namely that their algebraic reduct $\alg A$ admits only polynomially-many homomorphisms from $\alg X$ to $\alg A$ and, moreover, the homomorphisms can be enumerated in polynomial time. This is in sharp contrast to purely relational structures, where ``most'' tractable cases do not have this property---even the condition that counting the number of homomorphisms is in P (characterized in~\cite{Bulatov-counting}) is much stronger  than polynomial-time solvability of the CSP. 

More precisely, we introduce a class $\Kpoly$ of those algebras $\alg A$ for which there exists a polynomial $p$ such the number of homomorphisms from any finite $\alg X$ (of the same signature) into $\alg A$ is at most $p(|X|)$, where $X$ is the domain of $\alg X$. The class $\Kpolyeff$ is introduced by additionally demanding that all the homomorphisms can be enumerated in polynomial time. We currently do not have any example witnessing that $\Kpolyeff$ is a proper subset of $\Kpoly$.

Notice that every structure whose algebraic reduct is in $\Kpolyeff$ has its CSP in polynomial time, and this is the case for all the structures in the second item of Theorem~\ref{cor:boolean-dichotomy}. Our main technical contribution is a sufficient condition for membership in $\Kpolyeff$ which goes well beyond the Boolean case. Its precise formulation (Theorem~\ref{thm:membership-separating-polynomials}) requires some concepts from  Tame Congruence Theory~\cite{Hobby:1988} and here we only state a consequence.

\begin{restatable}{theorem}{idempotenttaylor}
\label{thm:idempotent-Taylor}
Every finite idempotent equationally nontrivial algebra $\rel A$ is in $\Kpolyeff$. 
\end{restatable}

In this theorem, \emph{idempotent} means that the identity $f(x, \dots, x) \approx x$ is satisfied for every operation, and \emph{equationally nontrivial} means that the term operations satisfy some identities that are not satisfiable by term operations in every algebra. For example, any idempotent algebra with a commutative term operation is equationally nontrivial; on the other hand, an algebra with an associative term operation may (but need not) be equationally trivial, because every algebra has an associative term operation, the projection onto the first variable.

The class of equationally nontrivial algebras has a prominent role in the classical CSP over relational structures: If $\rel A$ is a finite relational structure containing all the singleton unary relations (an assumption we can make without loss of generality in the relational setting) and $\alg A$ is the algebra of polymorphisms of $\rel A$, then $\alg A$ is idempotent and either it is equationally trivial (in which case $\csp(\rel A)$ is NP-complete~\cite{Bulatov:2005}), or it is equationally nontrivial (in which case $\csp(\rel A)$ is in P~\cite{Bulatov:2017, Zhuk:2017,Zhuk:2020}). 

We also complement Theorem~\ref{thm:idempotent-Taylor} with some other membership and non-membership results and obtain a complete classification  of 3-element algebras in $\Kpolyeff$ (Theorem~\ref{thm:3-element})  and algebras in $\Kpolyeff$ that are simple and contain all the constant operations (Theorem~\ref{thm:tame-algebras-characterization}).

Note that every expansion $\rel A$ of an algebra in $\Kpolyeff$ by relations or operations still has $\csp(\rel A)$ solvable in polynomial time. Thus members of $\Kpolyeff$ are \emph{inherently tractable} in this sense. It is consistent with the examples we have that $\Kpolyeff$ consists exactly of inherently tractable algebras.

Besides the significance of $\Kpolyeff$ in the complexity of CSP over general structures, we regard the problem of characterizing $\Kpoly$ and $\Kpolyeff$ as a very natural and fundamental problem in algebra,
that, to the best of our knowledge, has not appeared anywhere in the literature on this subject.

\subsection{Non-Collapse of Width}

A significant part of tractable CSPs over relational structures (e.g. HORN-SAT or 2-SAT) can be solved in polynomial-time by a very natural algorithm, which can be informally described as follows. We derive the strongest constraints on $k$-element subsets of $A$ (where $A$ is the domain of the instance $\rel A$) that can be derived by considering $l$-element subsets of $A$ at a time. Then we reject the instance if a contradiction is derived, otherwise we accept. If $1\leq k \leq l$ are fixed integers, this is a polynomial-time algorithm. If the answer of the algorithm is correct for every instance $\rel X$ of $\csp(\rel A)$, we say that $\rel A$ has \emph{width $(k,l)$}. We say that $\rel A$ has \emph{bounded width} if it has width $(k,l)$ for some $k,l$. There are various equivalent formulations of bounded width, e.g. in terms of descriptive complexity~\cite{Kolaitis:1995}.

Bounded width finite relational structures have been characterized in~\cite{Barto:2009c} and a refinement in~\cite{BartoCollapse} shows a surprising phenomenon: whenever $\rel A$ has bounded width, then it actually has relational width (2,3). 

It follows from our results that a lot of tractable CSPs over arbitrary Boolean structures have bounded width (even though it would be inefficient to use a $(k,l)$-algorithm to solve them). In particular, every CSP in the second item of Theorem~\ref{thm:boolean-dichotomy} has bounded width. However, the ``collapse of width'' phenomenon described above no longer occurs: we present two examples of structures that have bounded width, but do not have width (2,3). One is the two-element set with the \emph{Sheffer stroke} (i.e., the not-or operation), the other is the two-element set with the operations $(x,y)\mapsto x+y$ and $(x,y)\mapsto x+y+1$.

\begin{restatable}{theorem}{nocollapse}
\label{thm:no-collapse}
There are bounded width  finite structures that do not have relational width (2,3).    
\end{restatable}

These examples witness that CSPs over structures are substantially richer than CSPs over relational structures from the perspective of descriptive complexity.

\section{Basic definitions}\label{sect:basic-definitions}

We present here some elementary definitions.
Except for the notion of graph of a structure, all the terms defined here are standard.

\subsection{Structures and Algebras}
A \emph{signature} is a set $\sig S$ of function symbols and relation symbols, each of which has an \emph{arity}.
\begin{definition}[Structure]
Let $S$ be a signature.
An $S$-structure $\rel A$ consists of a set $A$, called the \emph{domain} of the structure, together with an assignment from $S$
to operations and relations on $A$  whose arities match the arities of the corresponding symbols in $\sig S$. These operations and relations are also referred to as \emph{basic operations and relations} of $\rel A$.
\end{definition}
A structure with a purely functional signature is called an \emph{algebra}; these will be denoted by bold letters. For a given structure, denoted by a bold or script letter, we typically denote the domain of that structure by the same letter in plain font.

If $f_1,\dots,f_n$ and $R_1,\dots,R_m$ are the symbols from $\sig S$, we write $\rel A=(A;f^{\rel A}_1,\dots,f^{\rel A}_n,R^{\rel A}_1,\dots,R^{\rel A}_m)$.
The \emph{algebraic reduct} of $\rel A$ is the structure $(A;f^{\rel A}_1,\dots,f^{\rel A}_n)$, whose signature only contains function symbols. Similarly, the \emph{relational reduct} of $\rel A$ is the relational structure $(A;R^{\rel A}_1,\dots,R^{\rel A}_m)$.

\begin{definition}[Graph of a structure]
Let $\rel A$ be an $S$-structure.
The \emph{graph of $\rel A$} is the structure $\graph(\rel A)$ with the same domain as $\rel A$,
whose signature $\sig S'$ contains all the relation symbols from $\sig S$ together with a $(k+1)$-ary relation symbol $G_f$ for every $f\in\sig S$ of arity $k$, which is interpreted in $\graph(\rel A)$ as $\{(a_1,\dots,a_k,b)\mid f^{\rel A}(a_1,\dots,a_k)=b\}$, while $R^{\graph(\rel A)}=R^{\rel A}$ for every relation symbol $R\in\sig S$.
\end{definition}

For \sig S-structures $\rel A, \rel B$, a \emph{homomorphism} from \rel A to \rel B is a map $h\colon A\to B$ such that
\begin{itemize}
\item for every function symbol $f\in\sig S$ of arity $k$ and every $a_1,\dots,a_k\in A$,
\[h(f^{\rel A}(a_1,\dots,a_k))=f^{\rel B}(h(a_1),\dots,h(a_k)), \text{ and}\]
\item for every relation symbol $R\in\sig S$ of arity $k$ and every $(a_1,\dots,a_k)\in R^{\rel A}$, we have $(h(a_1),\dots,h(a_k))\in R^{\rel B}$.
\end{itemize}
We simply write $h\colon\rel A\to\rel B$ to denote that $h$ is a homomorphim from $\rel A$ to $\rel B$,
and $\rel A\to\rel B$ for the statement that there exists a homomorphism from $\rel A$ to $\rel B$.
We denote by $\Hom(\rel A,\rel B)$ the set of homomomorphisms from $\rel A$ to $\rel B$.
A \emph{polymorphism} of a relational structure $\rel A$ is a map $h\colon A^n\to A$ such that for every relation $R^{\rel A}$ of $\rel A$ and all tuples $t_1,\dots,t_n\in R^{\rel A}$, the tuple $h(t_1,\dots,t_n)$, computed component-wise, is in $R^{\rel A}$.
We also say that the relations of $\rel A$ are \emph{preserved} by $h$.

\begin{definition}[Substructures]
A structure $\rel B$ is a \emph{substructure} of $\rel A$ if:
\begin{itemize}
    \item it has the same signature $S$ as $\rel A$,
    \item its domain is a subset of the domain of $\rel A$,
    \item $R^{\rel B} = R^{\rel A}\cap B^n$ for every relation symbol $R\in S$ of arity $n$ and $f^{\rel B} = f^{\rel A}|_{B}$ for every operation symbol $f\in S$.
\end{itemize}
In particular, the definition implies that the domain of $\rel B$ must be closed under every operation of $\rel A$. We say that $\rel B$ is a \emph{proper} substructure if its domain is strictly contained in the domain of $\rel A$.
\end{definition}

In case we deal with algebras, then we call $\alg B$ a \emph{subalgebra} of $\alg A$. 
The following concepts refines the standard notion of a subalgebra generated by a set, it will be convenient in some proofs.

\begin{definition}[Algebra generated by a set]
The algebra generated by a subset $B\subseteq A$ under some operations $f_1,\dots,f_k$ on $A$ is the algebra $(C;f_1|_C,\dots,f_k|_C)$ where $C$ is the minimal set containing $B$ and closed under the operations $f_1,\dots,f_k$ (the signature will be clear from the context).
\end{definition}

\subsection{Congruences}

\begin{definition}[Congruence]
Let $\theta$ be an equivalence relation on $A$.
We say that $\theta$ is a \emph{congruence} of a structure $\rel A$ with domain $A$ if every operation of $\rel A$ induces an operation on $A/\theta$, i.e., if the operations of $\rel A$ preserve $\theta$.
\end{definition}
For every structure $\rel A$, the equality relation, denoted by $0_A$, is a congruence of $\rel A$, and so is the full relation $A\times A$, denoted by $1_A$. A structure $\rel A$ is \emph{simple} if $|A|>1$ and it has only the trivial congruences $0_A$ and $1_A$.

Given an equivalence relation $\theta$ on $A$, we write $a/\theta$ to denote the $\theta$-equivalence class of $a$, that is, $a/\theta:=\{b\in X \mid (a,b)\in\theta\}$.

Given a structure $\rel A$ and a congruence $\theta$ of $\rel A$, we define the \emph{quotient structure} $\rel A/{\theta}$ as the structure with the same signature as $\rel A$, whose domain is the set $A/\theta$ of $\theta$-equivalence classes, and such that:
\begin{itemize}
    \item $f^{\rel A/\theta}(a_1/\theta,\dots,a_k/\theta) = f^{\rel A}(a_1,\dots,a_k)/\theta$ for every $k$-ary operation symbol $f$ in the signature of $\rel A$,
    \item $(a_1/\theta,\dots,a_k/\theta)\in R^{\rel A/\theta}$ iff there exist $b_1,\dots,b_k$ such that $(b_i,a_i)\in\theta$ for all $i$ and $(b_1,\dots,b_k)\in R^{\rel A}$, for every $k$-ary relation symbol $R$.
\end{itemize}

\subsection{Terms and identities}

Let $S$ be a signature.
An \emph{$S$-term}  $s$ over the set of variables $\{x_1 \dots, x_n\}$ is a finite rooted tree  whose leaves are labelled by variables $x_1,\dots,x_n$ (that do not necessarily all appear, and can appear more than once) and where the internal nodes are labelled by function symbols from $S$, in a way that the degree of every node matches the arity of the corresponding symbol. We sometimes write $s$ as $s(x_1, \dots, x_n)$ to specify the set of variables over which the term $s$ is considered.
If \rel A is an \sig S-structure and $s(x_1,\dots,x_n)$ an \sig S-term, then $s^\rel A\colon A^n\to A$ is the operation naturally induced by $s$ in $\rel A$. Such an operation is called a \emph{term operation} of $\rel A$.
A set of operations on a set $A$ is called a \emph{clone} if it is equal to the set of all term operations of an (algebraic) structure whose domain is $A$. The clone of term operations of a structure $\rel A$ is  denoted $\Clo{\rel A}$. Two structures $\rel A$ and $\rel B$ are \emph{term-equivalent} if $\Clo{\rel A} = \Clo{\rel B}$.

An \emph{$S$-identity} is a pair $(s,t)$ of $S$-terms over the same set of variables $\{x_1, \dots, x_n\}$, also denoted $s\approx t$. We write \(\rel A \models s\approx t\) to mean that \(s^{\rel A} = t^{\rel A}\) holds.
If $\Sigma$ is a collection of \sig S-identities, then $\rel A \models \Sigma$ means that \(\rel A \models s\approx t\) holds for all pairs \((s, t) \in \Sigma\).

An algebra $\alg A$ is a \emph{term-reduct} of an algebra $\alg B$ if $\Clo{\alg A} \subseteq \Clo{\alg B}$. 

\subsection{Constraint Satisfaction Problems}

For a fixed \sig S-structure \rel A, the \emph{constraint satisfaction problem} for \rel A, denoted $\csp(\rel A)$, is the following problem: given a finite \sig S-structure \rel X, decide whether there is a homomorphism from \rel X to \rel A. In this context, $\rel A$ is also called a \emph{template}. 

Note that a map $h\colon X\to A$ is a homomorphism $\rel X\to\rel A$ if, and only if, it is a homomorphism $\graph(\rel X)\to\graph(\rel A)$.
Therefore, $\csp(\rel A)$ admits a polynomial-time reduction to $\csp(\graph(\rel A))$, where the reduction maps an instance $\rel X$ of $\csp(\rel A)$ to the instance $\graph(\rel X)$ of $\csp(\graph(\rel A))$.
In particular, one can see $\csp(\rel A)$ as the problem $\csp(\graph(\rel A))$ where the instances are syntactically restricted, i.e., where some of the input relations are restricted to be graphs of operations.

There is in general no polynomial-time reduction from $\csp(\graph(\rel A))$ to $\csp(\rel A)$  unless $P=NP$, as we will see in Proposition~\ref{prop:3-element-example}.
However, Feder \emph{et al.} prove the following reduction from $\csp(\graph(\rel A))$ to $\csp(\rel A)$ when all the operations of $\rel A$ are essentially unary.

\begin{theorem}[\protect{\cite[Theorem 13]{Feder:2004}}]
  \label{thm:unary-ops-equivalence-relational}
	Let $\rel A$ be a structure whose operations are all essentially unary. Then $\homcsp{\rel A}$ and $\homcsp{\graph(\rel A)}$ are equivalent
	under polynomial-time many-one reductions.
\end{theorem}

Theorem~\ref{thm:unary-ops-equivalence-relational} does not extend to structures with operations of higher arity. In fact, Theorem~\ref{thm:boolean-dichotomy} implies that it fails for the two-element algebra $(\{0,1\},\cdot)$ with the single binary operation $x\cdot y:=\neg x\land\neg y$.

\section{Reductions}
\label{sect:general}

The first lemma shows that for any \emph{finite} set of identities $\Sigma$ satisfied by a template $\rel A$ we can efficiently enforce $\Sigma$ in any instance $\rel X$ of $\csp(\rel A)$ without changing the number of homomorphisms. We do not know whether the same is true for infinite sets of identities.

\begin{lemma}\label{lem:eq}
Let $\sig S$ be a signature and  \rel A be an \sig S-structure.
Let $\Sigma$ be a finite set of identities such that \(\rel A \models \Sigma\).
For every instance $\rel X$ of\/ $\csp(\rel A)$, one can compute in polynomial time an instance $\rel X'$ of\/ $\csp(\rel A)$ such that $\rel X'\models\Sigma$
and $\left|\Hom(\rel X,\rel A)\right|=\left|\Hom(\rel X', \rel A)\right|$.
\end{lemma}
\begin{proof}
	For each identity $s(x_1,\dots,x_n) \approx t(x_1,\dots,x_n)$ in $\Sigma$ and each tuple $\mbf a$ such that \(s^{\rel X}(\mbf a) \neq t^{\rel X}(\mbf a)\), one can compute the congruence $\theta$ generated by \(s^{\rel X}(\mbf a)\) and \(t^{\rel X}(\mbf a)\), that is, the smallest equivalence relation containing $(s^{\rel X}(\mbf a),t^{\rel X}(\mbf a))$ and that is preserved by all operations of $\rel X$. Let \(\rel X_1:= \rel X/\theta\), and note that \(|\rel X_1| < |\rel X|\).
	We show that there is a bijection from $\Hom(\rel X,\rel A)$ to $\Hom(\rel X_1,\rel A)$.
	For an arbitrary $h\in\Hom(\rel X,\rel A)$,
	we have that $h(s^{\rel X}(\mbf a))=s^{\rel A}(h(\mbf a))=t^{\rel A}(h(\mbf a))=h(t^{\rel X}(\mbf a))$ for every identity $s\approx t$ in $\Sigma$.
	Thus, $\theta$ is a subset of $\ker(h):=\{(x,y)\in X^2 \mid h(x)=h(y)\}$. 
	It follows that $h$ factors uniquely as the composition of homomorphisms $\rel X\xrightarrow{\pi}\rel X/{\theta}\xrightarrow{h'}\rel A$, where $\pi$ is the canonical projection map $x\mapsto x/\theta$.
	Thus, the map $h\mapsto h'$ is injective.
	It is additionally surjective, as any $g\colon\rel X/{\theta}\to\rel A$ is of the form $h'$, where $h:=g\circ\pi$.
	
	Iterating over all identities in $\Sigma$, possibly several times, at the final step we obtain a structure $\rel X_n$ that satisfies $\Sigma$ and is such that $\left|\Hom(\rel X,\rel A)\right|=\left|\Hom(\rel X_n,\rel A)\right|$.
Moreover, since the number of elements in the intermediate structures decreases at each step, $|\rel X_{i+1}| < |\rel X_i|$, the process finishes in polynomial time.
\end{proof}

The second lemma implies that we can expand a template by term operations and quantifier-free pp-definable relations without increasing the complexity. 

\begin{definition}[Quantifier-free pp-formula]
\label{def:qfpp}
Let $S$ be a signature. A \emph{quantifier-free pp-formula} is a conjunction of atomic $S$-formulas; i.e., a conjunctions of atoms $R(t_1,\dots,t_k)$ where $R$ is a relation symbol in $S$ and $t_1,\dots,t_k$ are $S$-terms.
\end{definition}

\noindent
The lemma is more general -- we can also take a term-reduct of the algebraic reduct of the template provided that the structures stay term-equivalent. 

\begin{lemma}\label{lem:general-reduction}
	Let \rel A and \(\rel B\) be term-equivalent structures, and suppose that every relation of \rel A has a quantifier-free pp-definition in \(\rel B\). Then $\homcsp{\rel A}$ has a polynomial-time many-one reduction to $\homcsp{\rel B}$.
\end{lemma}
\begin{proof}

Let $F=\{f_1, \dots, f_k\}$ be the set of operation symbols in the signature of $\rel A$ and $G=\{g_1, \dots, g_l\}$ be defined similarly for $\rel B$. 
In the following, if $s$ is an $F$-term and $t_1, \dots, t_k$ $G$-terms we write $s[t_1, \dots, t_k]$ for the $G$-term obtained by replacing every occurrence of the operation symbol $f_i$ in $s$ by the term $t_i$; similarly with roles of $F$ and $G$ swapped.

	By the fact that $\rel A$ and $\rel B$ are term-equivalent, there exist $G$-terms $t_1,\dots,t_k$ and $F$-terms $s_1,\dots,s_l$ such that
	$f^\rel A_i = t_i^{\rel B}$
	 for every $i\in\{1,\dots,k\}$ and 
	$g^\rel B_i = s_i^{\rel A}$
	 for every $i\in\{1,\dots,l\}$.
	Note that, as a  consequence, \rel A satisfies the identity
	$f_i \approx t_i[s_1, \dots,s_l]$.

	Let \rel X be an instance of $\csp(\rel A)$.
	By Lemma~\ref{lem:eq}, we can assume that \rel X also satisfies the identity $f_i \approx t_i[s_1, \dots,s_l]$ for each $i \leq k$.
	Let $\rel Y$ be the instance of $\csp(\rel B)$ with the same domain as $\rel X$ and whose relations and operations are defined as follows:
	\begin{itemize}
	\item For each operation symbol $g_i \in G$, let  $g_i^\rel Y = s_i^{\rel X}$.
	\item The relations of $\rel Y$ are defined as in the standard reduction between CSPs of relational structures:
	assuming $R^{\rel A}$ is $n$-ary and defined by a conjunction of atomic formulas of the form
	$T(r_1(x_1,\dots,x_n),\dots,r_m(x_1,\dots,x_n))$
	in $\rel B$, where $T$ is a relation symbol from the signature of $\rel B$ and each $r_{i}$ is a term in the signature of $\rel B$,
	every tuple $\mbf a$ in $R^{\rel X}$
	yields the corresponding tuple $(r^{\rel Y}_1(\mbf a),\dots,r^{\rel Y}_{m}(\mbf a))$ in $T^{\rel Y}$.
	\end{itemize}
	
	We prove that a map $h\colon X\to A$ is a homomorphism $\rel X\to\rel A$ iff it is a homomorphism $\rel Y\to\rel B$.
	For the relational constraints this follows from the known construction for CSPs of relational structures,
	so we only prove the statement for the operation constraints.
	Assume $h$ is a homomorphism $\rel X\to \rel A$. Consider any $g_i \in G$ and any tuple $\mbf x$ of the correct length.
	\begin{align*}
	g_i^\rel B(h(\mbf x)) &= s_i^{\rel A}(h(\mbf x)) & (\text{by definition of $s_i$})\\
	&= h(s_i^{\rel X}(\mbf x)) & (\text{since } h \colon \rel X\to\rel A)\\
	&= h(g_i^\rel Y(\mbf x)) & (\text{by definition of $\rel Y$}). 
	\end{align*}
	
	Suppose now that $h$ is a homomorphism $\rel Y\to\rel B$, $f_i \in F$ and  $\mbf x$ a tuple. Then
	\begin{align*}
	h(f_i^\rel X(\mbf x)) &= h(t_i[s_1, \dots, s_l]^{\rel X}(\mbf x))\\
	&= h(t_i^{\rel Y}(\mbf x))\\
	&= t_i^{\rel B}(h(\mbf x))\\
	&= f^\rel A_i(h(\mbf x))
	\end{align*}
	where the equalities follow (in order) from $\rel X$ satisfying $f_i \approx t_i[s_1, \dots, s_l]$, the definition of $\rel Y$, the fact that $h$ is a homomorphism $\rel Y\to\rel B$, and the definition of $t_i$.
	This concludes the proof.
\end{proof}

As a corollary of the previous lemma, we obtain an algebraic invariant of the complexity of the CSP of finite structures.
A partial operation $h\colon A^n\rightharpoonup A$ is a  \emph{partial polymorphism} of a relational structure $\rel A$ if for every relation $R$ of $\rel A$ and tuples $t_1,\dots,t_n\in R$, if $h(t_1,\dots,t_n)$ is defined, then it is in $R$.
It is known that if $\rel A$ and $\rel B$ are finite relational structures such that every partial polymorphism of $\rel B$ is a partial polymorphism of $\rel A$, then the relations of $\rel A$ all have a quantifier-free pp-definition in $\rel B$~\cite{PartialPoly}.

\invariant*

We note that in the Boolean case our result shows that membership in P or NP-hardness of the CSP depends only on the clone generated by the basic operations and on the (total) polymorphisms of $\graph(\rel A)$.
This invariant is \emph{weaker}, in the sense that it contains less information. Indeed, every polymorphism of $\graph(\rel A)$ is in particular a partial polymorphism of the relational reduct of $\rel A$.
It is unclear whether this weaker invariant is enough to separate tractable and NP-complete problems for structures with bigger domains.
We also do not know whether the term-equivalence in Lemma~\ref{lem:general-reduction} can be replaced by the inclusion $\Clo{\alg B} \subseteq \Clo{\alg A}$.

\section{Algebras with polynomially-many homomorphisms}

For an algebra $\alg A$, we denote by $\counting{\alg A}(n)$ the maximum value of $\left| \Hom(\alg X,\alg A)\right|$, where $\alg X$ ranges over algebras with $n$ elements.
Similarly, we let $\counting{\alg A}^s(n)$ be the sequence counting the maximum number of surjective homomorphisms from an $n$-element algebra $\alg X$ to $\alg A$.
We let $\Kpoly$ be the class of finite algebras $\alg A$ such that $\counting{\alg A}$ is bounded above by a polynomial, and similarly the class $\Ksurj$ of those finite algebras such that $\counting{\alg A}^s(n)$ is bounded above by a polynomial.
The effective variants of $\Kpoly$ and $\Ksurj$ are the classes $\Kpolyeff$ and $\Ksurjeff$ of algebras such that the corresponding homomorphisms can be enumerated in polynomial time.

We will be often working with polynomial operations and induced algebras and we now introduce the concepts. 

For an algebra $\alg A$ we denote by $\alg A^*$ the expansion of $\alg A$ by all the constant operations.
A \emph{polynomial operation} of an algebra $\alg A$ is  is a term operation of $\alg A^*$; in other words, an operation of the form $t(x_1,\dots,x_n,a_1,\dots,a_m)$ where $t$ is a term operation of $\alg A$ of arity $n+m$ and $a_1,\dots,a_m$ are elements from $A$. Two algebras are called \emph{polynomially equivalent} if they have the same universe and the same clone of polynomial operations.

For an algebra $\alg A$ (with universe $A$) and a subset $N \subseteq A$ we denote by $\alg A|_N$ the \emph{algebra induced by $\alg A$ on $N$}, that is, the algebra with universe $N$ whose operations are those polynomial operations of $\alg A$ that preserve $N$ (the signature of this algebra can be chosen arbitrarily so that the operation symbols are in bijective correspondence with the operations of $\alg A$).

The rest of this section is organized as follows. In Subsection~\ref{subsec:general} we prove several general results about classes $\Kpoly$ and $\Kpolyeff$, including Theorem~\ref{thm:membership-separating-polynomials}, which will serve as the main tool for membership results, 
together with the Tame Congruence Theory, whose basics are introduced in Subsection~\ref{subsec:tct}.
This is followed by two subsections providing sufficient conditions (Subsection~\ref{subsec:suff}) and necessary conditions (Subsection~\ref{subsec:nec}) for membership in these classes. Finally, in Subsection~\ref{subsec:thm} we state several corollaries of these conditions and give examples.
In particular, we classify the simple algebras in $\Ksurj$ (Theorem~\ref{thm:tame-algebras-characterization}), as well as the 3-element algebras (Theorem~\ref{thm:3-element}).
Finally, we use our results to classify the complexity of CSPs over Boolean structures (Theorem~\ref{thm:boolean-dichotomy}).

\subsection{General results} \label{subsec:general}

We start with some simple observations.
\begin{lemma}\label{lem:closure-properties}
    $\Kpoly$ and $\Kpolyeff$ are closed under finite products (of algebras with the same signature), subalgebras, and expansions.
    $\Ksurj$ and $\Ksurjeff$ are closed under finite products  and expansions.
\end{lemma}
However, we will see in Example~\ref{ex:not-closed-quotient} below that $\Kpoly$ is not closed under quotients. 

It follows from the next lemma that membership in any of the classes   depends only on the clone of operations of the algebra.
\begin{lemma}\label{lem:clone-Kpoly}
    Let $\mathcal K$ be one of $\Kpoly,\Ksurj,\Kpolyeff,\Ksurjeff$.
    Let $\alg A, \alg B$ be algebras of finite signatures such that $\Clo(\alg A) \subseteq \Clo(\alg B)$.
    If $\alg A\in\mathcal K$, so is $\alg B$.
\end{lemma}

The following lemma explains the relationship between $\Kpoly$ and $\Ksurj$.
\begin{lemma}\label{lem:equivalences-Kpoly}
Let $\alg A$ be a finite algebra.
The following are equivalent:
\begin{enumerate}
    \item $\alg A$ is in $\Kpoly$,
    \item Every subalgebra $\alg B$ of $\alg A$ (including $\alg A$) is in $\Ksurj$,
    \item $\alg A$ is in $\Ksurj$ and every proper subalgebra $\alg B$ of $\alg A$ is in $\Kpoly$.
\end{enumerate}
The same equivalences hold if $\Kpoly$ and $\Ksurj$ are replaced by their effective variants.
\end{lemma}
\begin{proof}
    (1) implies (2). This follows from the fact that every surjective homomorphism $\alg X\to\alg B$ is in particular a homomorphism $\alg X\to\alg A$, giving $\counting{\alg B}^s\leq\counting{\alg A}$.
    
    (2) implies (3). Every homomorphism $\alg X\to\alg B$ is a surjective homomorphism onto a subalgebra $\alg B'$ of $\alg B$. In particular, $\counting{\alg B}\leq \sum_{\alg B'\leq\alg B} \counting{\alg B'}^s\leq\sum_{\alg B<\alg A}\counting{\alg B}^s + \counting{\alg A}^s$, which is polynomially bounded by assumption.
    
    (3) implies (1). We have $\counting{\alg A}=\sum_{\alg B<\alg A}\counting{\alg B}^s +\counting{\alg A}^s$.
    Since the terms in the sum are all polynomially bounded, so is $\counting{\alg A}$.
\end{proof}

Moreover, whether a finite algebra belongs to $\Ksurj$ only depends on its polynomial operations.

\begin{lemma}\label{lem:Ksurj-polynomials}
    A finite algebra $\alg A$ is in $\Ksurj$ if, and only if, $\alg A^*$ is in $\Ksurj$.
    The same holds with $\Ksurjeff$ in place of $\Ksurj$.
\end{lemma}
\begin{proof}
    Since $\alg A*$ is an expansion of $\alg A$, we have $\counting{\alg A^*}^s\leq\counting{\alg A}^s$, which gives the ``only if'' direction.
    
    For the other direction, we prove that $\counting{\alg A}^s(n)\leq n^{|A|}\cdot\counting{\alg A^*}^s(n)$ for all $n\geq 1$.
    Let $A=\{a_1,\dots,a_k\}$.
    Fix $n$, and let $\alg X$ be an $n$-element algebra having $\counting{\alg A}^s(n)$ surjective homomorphisms to $\alg A$.
    By the pigeonhole principle, there are $x_1,\dots,x_k\in X$ such that at least $\counting{\alg A}^s(n)/{n^k}$ of these homomorphisms map $x_i$ to $a_i$.
    Therefore, letting $\alg X^*$ be the expansion of $\alg X$ by the constant maps associated with $x_1,\dots,x_k$,
    we obtain that $\counting{\alg A^*}^s(n)\geq \counting{\alg A}^s(n)/{n^k}$ and the result follows.
\end{proof}

The following theorem serves as the main tool for proving membership in $\Ksurj$ and $\Ksurjeff$. It roughly says that in order to prove that $\alg A$ is in $\Ksurj$ it is enough to find a congruence $\alpha$ such that $\alg A/\alpha$ is in $\Ksurj$ and, for each $\alpha$-block $B$, the polynomials mapping $B$ into an algebra in $\Ksurj$ separate points.

\begin{theorem}\label{thm:membership-separating-polynomials}
Let $\alg A$ be an algebra, and $\alpha$ be a congruence of $\alg A$ such that
\begin{itemize}
    \item $\alg A/\alpha$ is in $\Ksurj$, and
    \item for every $a\neq b$ with $a/\alpha=b/\alpha=:B$, there is a polynomial operation $f^{\alg A}$ of $\alg A$ such that $f^{\alg A}(a)\neq f^{\alg A}(b)$ and such that some finite reduct $\alg B_{a,b}$ of $\alg A|_{f^{\alg A}(B)}$ is in $\Ksurj$. 
\end{itemize}
Then $\alg A$ is in $\Ksurj$. The same holds with $\Ksurjeff$ in place of $\Ksurj$.
\end{theorem}
\begin{proof}
By Lemma~\ref{lem:Ksurj-polynomials}, we can assume that $\alg A$ has all the constant operations.
Moreover, by Lemma~\ref{lem:clone-Kpoly}, one can assume that for all $a\neq b$ with $a/\alpha=b/\alpha=B$ the operation $f_{a,b}$ from the statement as well as the (finite) set $F_{a,b}$ of operations witnessing that $\alg A|_{f_{a,b}(B)}$ is in $\Ksurj$ are also basic operations of $\alg A$.

Let $\alg X$ be a finite algebra and let $h\colon\alg X\to\alg A$ be a surjective homomorphism.
Note that $h$ induces a homomorphism $\tilde h\colon\alg X\to\alg A/{\alpha}$, which is obtained by composing $h$ with the canonical projection $\alg A\to\alg A/\alpha$.

For every $(a,b)\in\alpha$ with $a\neq b$, let $f^{\alg A}_{a,b}$ be the polynomial operation from the second item, and let $F^{\alg A}_{a,b}$ be the finite set of polynomial operations of $\alg B_{a,b}$.
We denote by $F^{\alg X}_{a,b}$ the set of polynomial operations obtained by interpreting the operations from $F^{\alg A}_{a,b}$ in $\alg X$.
Let $Y_{a,b}=\{x\in X\mid \exists y \text{ s.t. } x=f^{\alg X}_{a,b}(y) \text{ and } h(y)\in B\}$.
Consider the algebra $\alg Z_{a,b}$ generated by $Y_{a,b}$ under the operations $F^{\alg X}_{a,b}$.
The homomorphism $h$ induces by restriction a surjective homomorphism $h_{a,b}\colon \alg Y_{a,b}\to \alg B_{a,b}$.
Indeed, given any $x\in Y_{a,b}$, we have $x=f^{\alg X}_{a,b}(y)$ for some $y$ such that $h(y)\in B$ and therefore $
    h(x) =  h(f^{\alg X}_{a,b}(y))
    = f^{\alg A}_{a,b}(h(y))
    \in f^{\alg A}_{a,b}(B)$.
Thus, $h(Y_{a,b})\subseteq f^{\alg A}_{a,b}(B)$.
Since $f^{\alg A}_{a,b}(B)$ is closed under the operations from $F^{\alg A}_{a,b}$, it follows that $h(Z_{a,b})\subseteq f^{\alg A}_{a,b}(B)$.
Therefore the restriction of $h$ to $Z_{a,b}$ induces a homomorphism $h_{a,b}\colon\alg Z_{a,b}\to \alg B_{a,b}$.
The fact that $h_{a,b}$ is surjective can be checked similarly.

We prove that $h$ is uniquely determined by the collection of homomorphisms $\{h_{a,b}\mid (a,b)\in\alpha, a\neq b\}$ and $\tilde h$.
Indeed, suppose that $g$ is any homomorphism $\alg X\to\alg A$ such that $\tilde g=\tilde h$ and such that $g(c)\neq h(c)$ for some $c\in X$.
Since $\tilde g=\tilde h$, we have $(g(c),h(c))\in\alpha$.
Moreover:
\begin{align*}
    g(f^{\alg X}_{h(c),g(c)}(c))&=f^{\alg A}_{h(c),g(c)}(g(c))\\
    &\neq f^{\alg A}_{h(c),g(c)}(h(c))\\
    &=h(f^{\alg X}_{h(c),g(c)}(c)).
\end{align*}
Thus, $g_{g(c),h(c)}\neq h_{g(c),h(c)}$.

Thus, every surjective homomorphism $h\colon \alg X\to\alg A$ is uniquely determined by:
\begin{itemize}
    \item a surjective homomorphism $\tilde h\colon\alg X\to\alg A/{\alpha}$,
    \item at most $|A|^2$ surjective homomorphisms from algebras of size at most $|X|$ (themselves determined by $\tilde h$) to some algebras in $\Ksurj$.
\end{itemize}
Since the number of such homomorphisms is polynomially bounded, we obtain a polynomial bound for the number of surjective homomorphisms $\alg X\to\alg A$.

Moreover, assuming that the surjective homomorphisms to $\alg A/\alpha$ can be enumerated in polynomial time, as well as the homomorphisms $\alg Y_{a,b}\to\alg B_{a,b}$, then the homomorphisms $\alg X\to \alg A$ can be enumerated in polynomial time.
\end{proof}

\subsection{Tame Congruence Theory}\label{subsec:tct}

By Lemma~\ref{lem:Ksurj-polynomials},
whether $\alg A$ is in $\Ksurj$ only depends on the clone of polynomial operations in $\alg A$.
\emph{Tame congruence theory} initiated in~\cite{Hobby:1988} is a  theory of finite algebras especially targeted at polynomial operations. Its basic facts and concepts are essential for the results in this paper. This subsection reviews the prerequisities following mostly Chapter 2 of~\cite{Hobby:1988}. 

One of the core components of the theory is the classification of so-called minimal algebras.
An algebra $\alg M$ is \emph{minimal} if every unary polynomial operation of $\alg M$ is either a constant or a permutation.

\begin{theorem}[\cite{Palfy}]\label{thm:palfy}
    Let $\alg M$ be a finite minimal algebra. Then $\alg M$ is polynomially equivalent to one of the following:
    \begin{enumerate}
        \item An algebra with only unary operations, all of which are permutations.
        \item A vector space over a finite field.
        \item A 2-element boolean algebra.
        \item A 2-element lattice.
        \item A 2-element semilattice, e.g. the algebra $(\{0,1\};\wedge)$ where $\wedge$ is the binary maximum operation.
    \end{enumerate}
\end{theorem}

\noindent
A minimal algebra is said to have type $i$, for $i\in\{1,\dots,5\}$, if it is polynomially equivalent to an algebra appearing in the $i$th item of Theorem~\ref{thm:palfy}. 

Consider now an arbitrary finite algebra $\alg A$, not necessarily minimal. 
A pair $(\alpha,\beta)$ of congruences of $\alg A$ is called a \emph{cover} if $\alpha\subsetneq \beta$ and no congruence $\gamma$ of $\alg A$ lies strictly between $\alpha$ and $\beta$.
By using a construction explained below, for each cover $(\alpha,\beta)$ one obtains a minimal algebra that belongs to one of the five classes in the previous theorem and can therefore be given a type $i\in\{1,\dots,5\}$.
We then assign this type to the pair $(\alpha,\beta)$,
which we write $\typ(\alpha,\beta)=i$.

The construction goes as follows. Let $U$ be a minimal set with the property that there exists a unary polynomial $p$ of $\alg A$ such that $p(A)=U$ and $p(\beta)\not\subseteq\alpha$ (in the most important case for us, $\alpha=0_A$, this precisely means that $U$ intersects one $\beta$-block in at least 2 elements).
Then, for every $a\in U$ such that $a/\beta\cap U\not\subseteq a/\alpha$,
the induced algebra $\alg A|_N$ on the set $N:=(a/\beta\cap U)$ is such that $(\alg A|_N)/\alpha$ is a minimal algebra that is of one of the 5 types given in Theorem~\ref{thm:palfy}. 
Such a set $N$ is called an \emph{$(\alpha,\beta)$-trace}.
Under the assumption that $(\alpha,\beta)$ is a cover,\footnote{Most of the results we present hold in  the more general setting where the pair $(\alpha,\beta)$ satisfies a weaker condition than being a cover, called \emph{tameness}.
For simplicity, we refrain from presenting the results in this more general framework.} it can be shown that all the $(\alpha,\beta)$-traces have the same type independent on the chosen $U$ and $N$, and therefore $\typ(\alpha,\beta)$ is well-defined. For a simple algebra $\alg A$ we write $\typ(\alg A) := \typ(0_A,1_A)$ and talk about the type of $\alg A$. 

The following separation property is crucial for our applications of Theorem~\ref{thm:membership-separating-polynomials}.

\begin{theorem}[Corollary of Theorem~2.8(4) in~\cite{Hobby:1988}]\label{thm:minimal-congruence}
Let $\alg A$ be a finite algebra and $\alpha$ be a minimal  congruence distinct from $0_A$.
Then for every $a \neq b$ in $A$  with $a/\alpha = b/\alpha =: B$ there is a polynomial operation $f$ of $\alg A$ such that $f(a)\neq f(b)$ and
$f(B) = N$ for some $(0_A,\alpha)$-trace $N$.
\end{theorem}
\begin{proof}
Let $U$ be a minimal set with the property that there exists a unary polynomial operation $p$ of $\alg A$ such that $p(A)=U$ and $p(\alpha)\neq 0_A$.
Let $(a,b)\in\alpha$ with $a\neq b$.
By Theorem 2.18 in~\cite{Hobby:1988}, there exists a unary polynomial operation $f$ such that $f(a)\neq f(b)$ and $f(A)=U$; let $N$ be $f(a/\alpha)$.
Note that $N=f(a)/\alpha\cap U$ and that $N$ contains $f(a)\neq f(b)$, so that it is a $(0_A,\alpha)$-trace.
\end{proof}

\subsection{Membership}\label{subsec:suff}

We start this subsection by proving that every minimal algebra that is not of type $1$ belongs to $\Kpolyeff$. By Lemma~\ref{lem:clone-Kpoly}, it suffices to prove it for finite abelian groups and a two-element semilattice (since algebras of type 3 and 4 also have semilattice term operations).

\begin{lemma}\label{lem:semilattice-case}
	The semilattice $\alg S = (\{0, 1\}; \wedge)$ is in $\Kpolyeff$.
\end{lemma}
\begin{proof}
    Let $\alg X=(X;s^{\alg X})$ be a finite algebra in the same signature $\{s\}$. Consider the set of identities defining semilattices, that is, 
    $\Sigma =\{ s(x,x) \approx x, s(x,y) \approx s(y,x), s(s(x,y),z) \approx s(x,s(y,z))\}$
	By Lemma~\ref{lem:eq} we can assume that $s^{\alg X}$ is a semilattice operation.
	
	Define $x\leq y$ if $x = s^{\alg X}(x,y)$.
	It follows from the fact that $s^{\alg X}$ is a semilattice operation that $\leq$ is a partial order on $X$.
	For every homomorphism $h\colon\alg X\to\alg S$, the set $h^{-1}(\{1\})$ is a principal filter in $(X,\leq)$, that is,
	an upward closed set that is closed under $s^{\alg X}$.
	Indeed, if $x\in h^{-1}(\{1\})$ and $x\leq y$, then
	$h(y) = 1\wedge h(y) = h(x)\wedge h(y)=h(s^{\alg X}(x,y))=h(x)=1$.
	Similarly, if $x,y\in h^{-1}(\{1\})$, then $h(s^{\alg X}(x,y))=h(x)\wedge h(y)=1$.
	
	Since there are at most $|X|$ such principal filters, one obtains that there are at most $|X|$ homomorphisms from $\alg X$ to $\alg S$.
	All such homomorphisms can be enumerated can iterating over the elements $x$ of $X$, building the corresponding principal filter $x\!\!\uparrow\; :=\{y\in X\mid y\geq x\}$ and checking the map that sends $y$ to $1$ iff $y\in x\!\!\uparrow$ is a homomorphism.
\end{proof}

\begin{lemma}\label{lem:minority-case}
Let $\alg G$ be a finite group. Then $\alg G$ is in $\Kpolyeff$.
\end{lemma}
\begin{proof}
    Let $\alg X$ be a finite algebra in the same signature as $\alg G$.
	As above, we can assume by Lemma~\ref{lem:eq} that
	$\alg X$ is a group.
	
	Every homomorphism $\alg X\to\alg G$ is determined by its action on a generating set of $\alg X$.
	Since every group has a generating set of logarithmic size, there are at most $|G|^{\log|X|}= |X|^{\log|G|}=poly(|X|)$ homomorphisms from $\alg X$ to $\alg G$.
	Moreover, a generating set of $\alg X$ of logarithmic size can be computed greedily in polynomial time.
	Thus, one can enumerate all the homomorphisms from $\alg X$ to $\alg G$.
\end{proof}

The following theorem is our main membership result. It is obtained by combining the previous two lemmata, Theorem~\ref{thm:membership-separating-polynomials}, and Theorem~\ref{thm:minimal-congruence}. 

\begin{theorem}\label{thm:omit-type-1}
    Let $\alg A$ be a finite algebra with a sequence $\alpha_0,\dots,\alpha_n$ of congruences  such that
    \begin{itemize}
        \item $\alpha_0=0_A, \alpha_n=1_A$, and
        \item for all $i$, 
        $(\alpha_i,\alpha_{i+1})$ is a cover
        and
        $\typ(\alpha_i,\alpha_{i+1})\neq 1$.
    \end{itemize}
    Then $\alg A$ is in $\Ksurjeff$.
\end{theorem}
\begin{proof}
The proof is by induction on $n$.
    The case $n=0$ is clear, since then $|A|=1$.
    
    Suppose $n>0$. The algebra $\alg A/\alpha_1$ has the sequence of congruences $0_{A/\alpha_1}=\alpha_1/\alpha_1\subsetneq\alpha_2/\alpha_1\subsetneq\dots\subsetneq\alpha_n/\alpha_1=1_{A/\alpha_1}$,
    and $\typ(\alpha_i,\alpha_{i+1})=\typ(\alpha_i/{\alpha_1},\alpha_{i+1}/\alpha_i)$ (Corollary 5.3 in~\cite{Hobby:1988}).
    Therefore by induction we obtain that $\alg A/\alpha_1$ is in $\Ksurjeff$.
    Moreover, since $\typ(0_A,\alpha_1)\neq 1$, every $(0_A,\alpha_1)$-trace is in $\Kpoly$ by Lemma~\ref{lem:semilattice-case} and Lemma~\ref{lem:minority-case},
    and this fact is witnessed by a single binary polynomial operation.
     Then, by Theorem~\ref{thm:minimal-congruence} and Theorem~\ref{thm:membership-separating-polynomials},
     $\alg A$ is in $\Ksurjeff$.
\end{proof}

By Theorem~9.6 in~\cite{Hobby:1988}, if $\alg A$ is a finite equationally nontrivial, then no cover in any subalgebra of $\alg A$ has type 1 (the reason is that idempotent equational conditions are propagated to traces and idempotent reducts of minimal algebras of type 1 are equationally trivial). 
Theorem~\ref{thm:idempotent-Taylor} therefore follows from Theorem~\ref{thm:omit-type-1} and Lemma~\ref{lem:equivalences-Kpoly}(3) by induction on the size of $\alg A$. 

\idempotenttaylor*

For the classification of 3-element algebras in $\Kpoly$ we need an additional argument for the membership part.

\begin{lemma}\label{lem:3-element}
Let $\alg A$ be an algebra with three elements.
If every minimal congruence $\alpha$ of $\alg A$ is such that $\typ(0_A,\alpha)$ is not 1, then $\alg A$ is in $\Ksurjeff$.
\end{lemma}

\begin{proof}
We proceed by case distinction on the number of congruences $n\in\{2,3,4,5\}$ of $\alg A$.
If $n=2$ then $\alg A$ is simple, and the result follows from the previous theorem by taking $(\alpha_0,\alpha_1) := (0_A,1_A)$. 
If $n\in\{4,5\}$, then the fact that no cover $(0,\alpha)$ of congruences $\alg A$ has type 1 implies that none of the covers $(\alpha,1_A)$ has type 1 either (Lemma 6.2 in~\cite{Hobby:1988}).
Thus, the result follows from Theorem~\ref{thm:omit-type-1}.

It remains to consider the case where $\alg A$ has exactly three congruences, i.e., the congruence lattice of $\alg A$ is the 3-element chain $0_A\subsetneq\alpha\subsetneq 1_A$, where up to relabelling of the elements, $A=\{0,1,2\}$ and $\alpha$ is the congruence $\{0,1\}^2\cup\{(2,2)\}$.
Let $U$ be a minimal set with the property that there exists a unary polynomial $p$ of $\alg A$ such that $p(A)=U$ and $p(\alpha)\neq 0_A$.
Since $p(\alpha)\neq 0_A$, we have $0,1\in U$.

If $U\neq\{0,1,2\}$, then $U=\{0,1\}$ is a $(0_A,\alpha)$-trace, so that $\alg A|_U$ is in $\Ksurjeff$.
Moreover, the set
$\eta$ of pairs $(a,b)$ that \emph{cannot} be separated by a polynomial operation $q^{\alg A}$ such that $q^{\alg A}(A)=U$ is a congruence of $\alg A$.
By Theorem~\ref{thm:minimal-congruence},  $\alpha\cap \eta=0_A$. Given that $\alpha$ is the unique minimal congruence of $\alg A$, it follows that $\eta=0_A$. 
Thus, every pair $a,b$ with $a\neq b$ can be separated by a polynomial with range $U$.
By applying Theorem~\ref{thm:membership-separating-polynomials} (with the congruence $1_A$), we obtain $\alg A\in\Ksurjeff$.

Otherwise, $U=\{0,1,2\}$.
By the fact that $\typ(0_A,\alpha)$ is not 1, there is a binary polynomial $t^{\alg A}$ of $\alg A$ that acts on the $(0_A,\alpha)$-trace $N:=\{0,1\}$ as a semilattice operation or as addition modulo $2$. If $t^{\alg A}$ is a semilattice, one can assume it is $\wedge$ up to exchanging the roles of $0$ and $1$.
Then, regardless whether $t^{\alg A}$ is addition modulo $2$ or $\wedge$, the range of the unary polynomial operation $q^{\alg A}(x):=t^{\alg A}(1,x)$
contains $\{0,1\}$.
Since $\{0,1\}$ is not the image of a polynomial (by minimality of $U$), we obtain that $q^{\alg A}(2)=2$, so that $t^{\alg A}(1,2)=2$.
The symmetric argument shows that 
$t^{\alg A}(2,1)=2$.
Thus, $\alg A/\alpha$ is a two-element algebra in which $t$ induces  a binary commutative operation, and it follows that $\typ(\alg A/\alpha)=\typ(\alpha,1_A)$ is not 1. By Theorem~\ref{thm:omit-type-1}, this proves that $\alg A$ is in $\Ksurjeff$.
\end{proof}

\subsection{Non-membership}\label{subsec:nec}

Typical examples of algebras outside of $\Kpoly$ are algebras whose operations are unary.
We prove a generalization of this fact for finite algebras $\alg A$ whose term operations all have \emph{bounded essential arity}.
An $n$-ary operation $f\colon A^n\to A$ is \emph{essentially $k$-ary} if there exist $g\colon A^k\to A$ and $i_1,\dots,i_k\in\{1,\dots,n\}$ such that for all $\mbf a\in A^n$, the equality $f(a_1,\dots,a_n)=g(a_{i_1},\dots,a_{i_n})$ holds.

Our proof relies on the standard concept of \emph{free algebras}.
We denote the free algebra for an algebra $\alg A$ over generators $\{1,2, \dots, n\}$ by $\alg F(n)$. 
The following two properties are crucial for our purposes. First, $\alg F(n)$ is isomorphic to the subalgebra of $\alg A^{A^n}$ formed by the $n$-ary term operations of $\alg A$. In particular, $|F(n)| = |\Clo_n(\alg A)|$. Secondly, 
any mapping  $\phi\colon\{1,\dots,n\}\to A$ has a unique extension to a homomorphism $\alg F(n)\to\alg A$.
In particular, the number of surjective homomorphisms from $\alg F(n)$ to $\alg A$ is at least  the number of surjective maps from $\{1,\dots,n\}$ to $A$.
This number is the Stirling number of the second kind $\genfrac\{\}{0pt}{}{n}{|A|}$, which is asymptotically equivalent to $|A|^n$.

\begin{lemma}\label{lem:bounded-arity}
Let $\alg A$ be a finite algebra such that $\Clo(\alg A)$ has bounded essential arity.
Then $\counting{\alg A}^s(n)\geq 2^{n^{\Omega(1)}}$.
\end{lemma}
\begin{proof}
Suppose that $\Clo(\alg A)$ is essentially $k$-ary, which implies that $|\alg F(n)|=O(n^k)$.
Note that $\alg F(n)$ admits asymptotically $|A|^{n}$ surjective homomorphisms to $\alg A$, and $|A|^n \geq |A|^{\sqrt[k]{|\alg F(n)|}}=|A|^{|\alg F(n)|^{\Omega(1)}}$.
Thus, if $\alg A$ has bounded essential arity, then $\counting{\alg A}^s(n)\geq 2^{n^{\Omega(1)}}$.
\end{proof}

For our classification of 3-element algebras in $\Ksurj$, we need an additional fact proved below.
In particular, this shows that bounded essential arity does not characterize non-membership in $\Ksurj$.
\begin{lemma}\label{lem:exponentially-many-homos}
    Let $\alg A$ be an algebra on 3 elements.
    Suppose that $0_A\subsetneq\alpha\subsetneq 1_A$ is a congruence such that $\typ(0_A,\alpha)=1$.
    Then $\counting{\alg A}^s(n)\geq 2^{\Omega(n)}$.
\end{lemma}
\begin{proof}
    Without loss of generality, we let $A=\{0,1,2\}$ and $\alpha=\{0,1\}^2\cup\{(2,2)\}$.
    By assumption, every polynomial operation of $\alg A$ that preserves $\{0,1\}$ is essentially unary on $\{0,1\}$.
    
    Let $f^{\alg A}$ be an $n$-ary basic operation of $\alg A$ and let $I\subseteq\{1,\dots,n\}$.
    We claim that there exists a unary polynomial operation $q^{\alg A}$ and $i\in\{1,\dots,n\}$ such that:
    \begin{itemize}
        \item $f^{\alg A}(a_1,\dots,a_n)=q^{\alg A}(a_i)$
    holds for all $a_1,\dots,a_n$ such that $a_j=2$ iff $j\in I$.
        \item $q^{\alg A}$ is the constant operation with value $2$ or an operation preserving $\{0,1\}$, in which case $i\not\in I$,
    \end{itemize}
    Indeed, let $p^{\alg A}$ be the polynomial operation obtained from $f$ by fixing the $j$th argument to $2$, for every $j\in I$.
    In case $p^{\alg A}$ preserves $\{0,1\}$,
    then by assumption $p^{\alg A}$ is essentially unary, which gives $q^{\alg A}$ and $i$ as in the claim.
    So assume that $p^{\alg A}$ does not preserve $\{0,1\}$, i.e., for some tuple $\mbf a$ with entries in $\{0,1\}$, one has $p^{\alg A}(\mbf a)=2$.
    Then for \emph{every} tuple $\mbf b$ with entries in $\{0,1\}$, all the pairs $(a_i,b_i)$ are in $\alpha$, so that $(p^{\alg A}(\mbf a),p^{\alg A}(\mbf b))\in\alpha$. Since $p^{\alg A}(\mbf a)=2$, we obtain $p^{\alg A}(\mbf b)=2$.
    Then by choosing $q^{\alg A}$ to be the constant operation with value 2, and $i$ arbitrary, the claim is proved.

    Let $\alg F(n)$ be the free algebra for $\alg A|_{\{0,1\}}$ with $n$ generators.
    Note that since $\alg A|_{\{0,1\}}$ is essentially unary, $\alg F(n)$ has linear size in $n$.
    Adjoin a single element to $\alg F(n)$, that we denote by $\star$.
    Let $\alg X$ be the algebra in the signature of $\alg A$ defined on $X=F(n)\cup\{\star\}$ as follows. Let $f^{\alg A}$ be a basic operation of $\alg A$ of arity $n$, and let $\mbf x$ be an $n$-tuple.
    Let $I\subseteq\{1,\dots,n\}$ be the set of indices $j$ such that $x_j=\star$.
    Let $q^{\alg A}$ and $i$ be obtained from the claim above.
    If $q^{\alg A}$ is the constant operation equal to $2$, define $f^{\alg X}(\mbf x)=\star$.
    If $q^{\alg A}$ preserves $\{0,1\}$, then it is a basic operation of $\alg A|_{\{0,1\}}$ and $x_i\in F(n)$, and therefore $q^{\alg F(n)}(x_i)$ is well-defined.
    We let $f^{\alg X}(\mbf x)=q^{\alg F(n)}(x_i)$.
    
  We claim that there are at least $2^{\Omega(n)}$ surjective homomorphisms from $\alg X$ to $\alg A$.
    Let $\{1,\dots,n\}$ be the generators of $\alg F(n)$.
    By the definition of free algebras, every surjective map $\phi\colon\{1,\dots,n\}\to \{0,1\}$ extends to a surjective homomorphism $\Phi\colon\alg F(n)\to\alg A|_{\{0,1\}}$.
    We show that by additionally defining $\Phi(\star)=2$, we obtain a homomorphism $\alg X\to\alg A$.
    Let $f$ be an $n$-ary symbol and $\mbf x\in X^n$.
    Let $I\subseteq\{1,\dots,n\}$ be defined as above, and let $q^{\alg A}$ and $i$ be from the claim.
    If $q^{\alg A}$ is constant equal to $2$, then
    $f^{\alg A}(\Phi(x_1),\dots,\Phi(x_n)) = 2=\Phi(\star)=\Phi(f^{\alg X}(x_1,\dots,x_n))$.
    Otherwise, $f^{\alg A}(\Phi(x_1),\dots,\Phi(x_n))=q^{\alg A}(\Phi(x_i))=\Phi(q^{\alg F(n)}(x_i))=\Phi(f^{\alg X}(x_1,\dots,x_n))$.
    
    Thus, there are asymptotically $|A|^n$ surjective homomorphisms from $\alg X$ to $\alg A$, and $n=\Omega(|X|)$, which concludes the proof.
\end{proof}

Note that in Lemma~\ref{lem:bounded-arity} and Lemma~\ref{lem:exponentially-many-homos}, we proved $\alg A \not\in  \Kpoly$ by selecting a  $B \subseteq A$ and constructing, for each finite set $Y$, an algebra $\alg X$ whose domain contains $Y$ and has size $O(|Y|^k)$  such that every mapping $Y \to B$ extends to a homomorphism $\alg X \to \alg A$. We now observe that in such a situation, the algebra $\alg A$ is not inherently tractable, i.e., there exists an expansion of $\alg A$ with NP-complete CSP.
Indeed, pick any relation $R$ on $B$ such that $\csp(B;R)$ is NP-complete. Now $\csp(B;R)$ reduces to the CSP over $\alg A$ expanded by $R$  by the reduction that maps an instance $(Y;S)$ to the expansion of $\alg X$ by the relation $S$.

\subsection{Consequences and examples}\label{subsec:thm}

\begin{theorem}\label{thm:tame-algebras-characterization}
Let $\alg A$ be a finite simple algebra.
Then the following are equivalent:
\begin{enumerate}
    \item $\alg A$ is in $\Ksurjeff$ (equivalently, $\alg A^* \in \Kpolyeff$),
    \item $\alg A$ is in $\Ksurj$ (equivalently, $\alg A^* \in \Kpoly$),
    \item $\typ(\alg A)\neq 1$.
\end{enumerate}
\end{theorem}
\begin{proof}
The implication from 1) to 2) is trivial.
The implication from 2) to 3) is by contrapositive. By Theorem 13.3 in~\cite{Hobby:1988}, $\alg A$ is a so-called \emph{subreduct of the matrix product} $\alg N^{[k]}$, where $\alg N$ is an algebra with only constant (unary) operations.
Concretely, the universe of $\alg A$ is a subset of $N^k$ and each term operation $f$ of $\alg A$, say of arity $m$, is of the form
$$
f^{\alg A}(\mbf a_1, \dots, \mbf a_m) = (u_1(a_{i_1,j_1}), u_2(a_{i_2,j_2}), \dots, u_k(a_{i_k,j_k})),
$$
for some $1 \leq i_1, \dots, i_k \leq m$, $1 \leq j_1, \dots, j_k \leq k$, and unary operations $u_1, \dots, u_k: N \to N$.
In particular, one sees that every term operation of $\alg A$ is essentially $k$-ary.
By Lemma~\ref{lem:bounded-arity}, it follows that $\alg A$ is not in $\Ksurj$.

The implication from 3) to 1) follows from  Theorem~\ref{thm:minimal-congruence} and Theorem~\ref{thm:membership-separating-polynomials}, in the case that $\alpha$ is the full congruence.
\end{proof}

Our classification for 3-element algebras follows from Lemma~\ref{lem:3-element} and Lemma~\ref{lem:exponentially-many-homos}.
\begin{theorem}\label{thm:3-element}
Let $\alg A$ be an algebra over 3 elements.
Then $\alg A$ is in $\Ksurj$ if, and only if, every minimal congruence $\alpha$ of $\alg A$ is such that $\typ(0_A,\alpha)$ is not 1.
Moreover, if $\alg A$ is in $\Ksurj$, then it is in $\Ksurjeff$.
\end{theorem}

In particular, we obtain the following example of an algebra $\alg A$ in $\Kpoly$ that has a quotient $\alg A/\alpha$ that is not in $\Kpoly$.

\begin{example}\label{ex:not-closed-quotient}
Let $\alg A$ be the algebra on $\{0,1,2\}$ with %
the binary operation $\circ$ defined by the following table:

\[
\begin{array}{r|ccc}
\circ & 0 & 1 & 2\\ 
\hline
0 & 0 & 1 & 2\\
1 & 1 & 1 & 2 \\
2 & 1 & 0 & 2
\end{array}
\]

Let $\alpha$ be $\{0,1\}^2\cup \{(2,2)\}$. One can see that $\alpha$ is the only non-trivial congruence of $\alg A$ and that $\alg A/\alpha$ is an essentially unary 2-element algebra.
By Theorem~\ref{thm:tame-algebras-characterization}, $\alg A/\alpha$ is not in $\Kpoly$.
However, note that $\{0,1\}$ is the image of the polynomial $p(x):= x\circ 0$, thus it is a $(0_A,\alpha)$-trace on which $\circ$ induces a semilattice operation.
Therefore, $\typ(0_A,\alpha)\neq 1$ and $\alg A$ is in $\Ksurj$ by Theorem~\ref{thm:3-element}.
Since all the subalgebras of $\alg A$ are in $\Kpoly$, it follows from Lemma~\ref{lem:equivalences-Kpoly} that $\alg A$ is in $\Kpoly$.

It might help to explain how the proof of Theorem~\ref{thm:membership-separating-polynomials} works for this example.
Consider the polynomials $p(x):=x\circ 0$ (which serve as $p_{0,2}$ and $p_{0,1}$ in the proof of Theorem~\ref{thm:membership-separating-polynomials}) and $q(x):=x\circ 1$ (which serves as $p_{1,2}$).
When applied to $0,1,2$, these polynomials take the following values:
\[\begin{array}{c|c|c}
     x & p(x) & q(x) \\
      \hline
    0 & 0   & 1\\
    1 & 1 & 1\\
    2 & 1 & 0
\end{array}\]
We see that no two rows $(p(x),q(x))$  are identical, i.e., given $(p(x),q(x))$, $x$ can be recovered uniquely.
Let $\alg X$ be an algebra, and let $X_1:=p(X)$ and $X_2:=q(X)$.
Given two maps $h_1\colon X_1\to \{0,1\}$ and $h_2\colon X_2\to\{0,1\}$, there is at most one way that we can obtain a homomorphism $h\colon\alg X\to\alg A$ that extends $h_1$ and $h_2$, because every $h(x)$ is uniquely determined by $h(p(x))=h_1(p(x))$ and $h(q(x))=h_2(q(x))$.
Since $\alg N=(\{0,1\},\circ)$ is a semilattice, it follows from Lemma~\ref{lem:semilattice-case} that $\alg N$ is in $\Kpolyeff$.
Thus, there are only polynomially many homomorphisms from the algebra generated by $X_1$ under $\circ^{\alg X}$ to $\alg N$, and similarly for the algebra generated by $X_2$ under $\circ^{\alg X}$.
\end{example}

\booleandichotomythm*

\begin{proof}[Proof of Theorem~\ref{thm:boolean-dichotomy}]
Let $\rel A$ be a Boolean structure, and suppose that $\rel A$ has an operation that is not essentially unary.
Let $\alg A$ be the algebraic reduct of $\rel A$, and note that it is a simple  algebra.
If $\alg A$ does not have type 1, then it is in $\Kpolyeff$ and $\csp(\rel A)$ is in P.
Otherwise it has type 1 and all its operations are essentially unary.
Then $\csp(\rel A)$ and $\csp(\graph(\rel A))$ are polynomial-time equivalent by Theorem~\ref{thm:unary-ops-equivalence-relational}.
\end{proof}

By Theorem~\ref{thm:boolean-dichotomy}, the two sources of tractability in the boolean case are the following:
\begin{itemize}
    \item The graph of the structure is solvable in polynomial time,
    \item The algebraic reduct of the structure is in $\Kpolyeff$.
\end{itemize}

We give an example showing that in general, tractability does not come from either of these two sources.
\begin{proposition}\label{prop:3-element-example}
    There exists a 3-element structure $\rel A$ such that:
    \begin{itemize}
        \item $\csp(\rel A)$ is in P,
        \item $\csp(\graph(\rel A))$ is NP-complete,
        \item the algebraic reduct $\alg A$ of $\rel A$ is not in $\Kpolyeff$.
    \end{itemize}
\end{proposition}
\begin{proof}
    Let $\rel A$ be the structure over $\{0,1,2\}$, with one binary relation $E:=\{(x,y)\mid x\neq y\}$ and one binary operation $f$ defined by the following table:
    \[ \begin{array}{c|ccc}
        f & 0 & 1 & 2\\
         \hline
        0 & 0 & 1 & 2\\
        1 & 0 & 1 & 2\\
        2 & 2 & 1 & 2
    \end{array}\]
    Clearly $\csp(\graph(\rel A))$ is NP-hard, since  $\csp(\{0,1,2\},E)$ is the 3-coloring problem.
    Moreover, $\alg A$ is not in $\Kpoly$ since the subalgebra induced by $\{0,1\}$ is not.
    
    It can be seen that $\alg A$ has the single congruence $\beta=\{0,2\}^2\cup\{(1,1)\}$,
    and $\typ(0,\beta)\neq 1$.
    By Theorem~\ref{thm:3-element}, we obtain that $\alg A\in\Ksurjeff$.
    
    The following algorithm taking as input $\rel X$ solves $\csp(\rel A)$ in polynomial time:
    \begin{itemize}
        \item Let $\alg X$ be the algebraic reduct of $\rel X$.
        Since $\alg A\in\Ksurjeff$, one can first enumerate all the surjective homomorphisms $\alg X\to\alg A$ and check whether any of them is a homomorphism $\rel X\to \rel A$.
        \item Otherwise, one enumerates the homomorphisms $\alg X\to\alg A$
        whose image is a subset of $\{0,2\}$.
        This is possible, since the algebra induced by $\{0,2\}$ is a semilattice,
        and is in $\Kpolyeff$ by Lemma~\ref{lem:semilattice-case}.
        \item Supposing that no homomorphism has been found so far, one checks for homomorphisms $\alg X\to\alg A$ whose image is either included in $\{0,1\}$ or in $\{1,2\}$ as follows. Let $\rel B$ be the substructure of $\rel A$ induced by the set $\{0,1\}$.
        We have that $\csp(\rel B)$ reduces to $\csp(\graph(\rel B))$, and $\graph(\rel B)$ has a majority polymorphism.
        Therefore, one can decide the existence of a homomorphism $\rel X\to\rel B$ in polynomial time.
        One can repeat this procedure to check for a homomorphism $\rel X\to\rel C$, where $\rel C$ is the substructure of $\rel A$ induced by the set $\{1,2\}$.
    \end{itemize}
\end{proof}

\section{Bounded Relational Width}

We study here the notion of \emph{relational width}, as defined in~\cite{BartoCollapse}.

\begin{definition}[$k$-forth property and $(k,l)$-system]
Let $X,A$ be sets.
Let $C\subseteq A^L$, where $L\subseteq X$.
Let $k$ be an integer.
Let $\mathcal P=(P_K)_{K\subseteq X, |K|\leq k}$ be a family of sets, where $P_K\subseteq A^K$.
We say that $\mathcal P$ has the $k$-forth property for $C$ if for every $K\subseteq L$ such that $|K|\leq k$, and every $f\in P_K$,
there exists $g\in C$ such that $g|_K = f$ and for all $K'\subseteq L$ with $|K'|\leq k$, $g|_{K'}\in P_{K'}$.

Let $l\geq k$.
We say that $\mathcal P$ is a \emph{$(k,l)$-system} from $X$ to $A$ if it has the $k$-forth property for $A^L$, for every $L\subseteq X$ with $|L|\leq l$.
Such a system is \emph{non-trivial} if no $P_K$ is empty.
\end{definition}

\begin{definition}[Compatible system]
Let $\rel X$ and $\rel A$ be structures, and let $\mathcal P$ be a $(k,l)$-system from $X$ to $A$.
We say that $\mathcal P$ is \emph{compatible} with $\rel X$
if $\mathcal P$ additionally has the $k$-forth property for
\[ \{ g\colon L\to A \mid (g(y_1),\dots,g(y_n))\in R^{\graph(\rel A)}\}\]
for every relation $R$ and every $L=\{y_1,\dots,y_n\}$ such that $(y_1,\dots,y_n)\in R^{\graph(\rel X)}$.
\end{definition}

We say that $\rel A$ has \emph{relational width $(k,l)$} if every instance $\rel X$ of $\csp(\rel A)$ that has a non-trivial compatible $(k,l)$-system has a homomorphism to $\rel A$.
We say that $\rel A$ has \emph{bounded relational width} if it has relational width $(k,l)$ for some $k,l$.

It is known that it is possible to determine in polynomial time whether an instance $\rel X$ has a non-trivial compatible $(k,l)$-system. Indeed, the $(k,l)$-minimality algorithm~\cite{BartoCollapse} runs in polynomial time and returns the largest such system (with respect to inclusion of sets).\footnote{One could equivalently define ``having relational width $(k,l)$'' as ``the $(k,l)$-minimality algorithm correctly solves the CSP''; for our purposes, it is better to work with the present definition instead. The fact that the two definitions are equivalent is proved in~\cite{BartoCollapse}.}
Thus, CSPs with bounded relational width can in particular be solved in polynomial time.

For CSPs of finite relational structures, a collapse of the bounded relational width hierarchy occurs:
\begin{theorem}[\cite{BartoCollapse}]\label{thm:bw-collapse}
	Let $\rel A$ be a finite relational structure that has relational width $(k,l)$ for some $1\leq k\leq l$.
	Then $\rel A$ has relational width $(2,3)$.
\end{theorem}

We show here that this collapse does not occur for arbitrary finite structures, already over the two-element domain.

Let $\rel A = (\{0, 1\}; \cdot)$ be the structure %
where $x \cdot y = \neg x \land \neg y$.
We show that $\rel A$ has bounded relational width but does not have relational width $(2,l)$ for any $l$.
In the next two proofs, we use the notation $\rel A[X]$, where $\rel A$ is a relational structure and $X\subseteq A$, to denote the substructure induced by $X$ in $\rel A$.
\begin{proposition}\label{prop:not-23}
	$\rel A$ does not have relational width $(2,l)$, for any $l\geq 2$.
\end{proposition}
\begin{proof}
	Let $\rel X = (X; *)$ be the structure with domain $X = \{0, a_0, a_1, a_2, a_3, \bar 0, \bar a_0, \bar a_1, \bar a_2, \bar a_3\}$ and binary operation $*$ defined as follows: for all \(x \in X\) and all \(i \neq j\),
	\[
\begin{gathered}
          x * 0 = \bar x = 0 * x, \quad x * \bar 0 = 0 = \bar 0 * x, \quad x * \bar x = 0 = \bar x * x,\\
          x * x = \bar x, \quad \bar x * \bar x = x,\quad
          a_0 * a_1 = a_2,\\ \quad a_1*a_0 = a_3,\quad a_2 * a_3 = a_0, \quad a_3 * a_2 = a_1,\\
         \bar a_i * a_j = a_i = a_j * \bar a_i, \quad \bar a_i * \bar a_j = 0, 
\end{gathered}
\]
and for pairs where $*$ is not yet defined, let \(a_i * a_j = a_k\) for $k\not\in\{i,j\}$ in an arbitrary way.

Note that there is no homomorphism from $\rel X$ to $\rel A$. To see this, suppose the contrary and let \(f : \rel X \to \rel A\) be such a homomorphism.  Then we must have \(f(a_2) = f(a_3), f(a_0)=f(a_1)\). If both \(a_2\) and \(a_3\) are mapped to 1, then for all \(x\), we have \(a_2 * x \neq a_3\), but this contradicts \(a_2 * \bar a_3 = a_3\).  If both are mapped to 0, then $a_0$ and $a_1$ are both mapped to 1. Then again, this implies $0=f(a_0)\cdot f(\bar a_1)=f(a_0*\bar a_1)=f(a_1)=1$, a contradiction.

We now exhibit the following non-trivial $(2,l)$-system $\mathcal P$ that is compatible with $\rel X$:
\begin{itemize}
\item $P_{\{0,\bar 0\}} = \{ 0\mapsto 0,\bar 0 \mapsto 1\}$,
\item $P_{K} = \{f\colon K\to\{0,1\}\mid f(0)=0\}$ for $K$ containing $0$,%
\item $P_{K} = \{f\colon K\to\{0,1\}\mid f(\bar 0)=1\}$ for $K$ containing $\bar 0$,
\item $P_{\{a_i,a_j\}} = \{f\colon\{a_i,a_j\}\to\{0,1\}\mid (f(a_i),f(a_j))\neq(1,1)\}$, %
\item $P_{\{a_i,\bar a_j\}} = \{f\colon\{a_i,\bar a_j\}\to\{0,1\}\mid (f(a_i),f(\bar a_j))\neq(1,0)\}$,%
\item $P_{\{\bar a_i,\bar a_j\}} = \{f\colon\{\bar a_i,\bar a_j\}\to\{0,1\}\mid (f(\bar a_i),f(\bar a_j))\neq(0,0)\}$,%
\item $P_{\{a_i,\bar a_i\}} = \{f\colon\{a_i,\bar a_i\}\to\{0,1\}\mid \text{$f$ not constant}\}$,%
\end{itemize}
This is indeed a $(2,l)$-system: for every $K$ and every $f\in P_K$, it can be seen that there exists $g\colon X\to\{0,1\}$ whose restriction to every $2$-element subset $K'$ is in $P_{K'}$.
Moreover $\mathcal P$ is compatible with $\rel X$,
since every $f\in P_K$ can be extended to a partial homomorphism $g\colon \graph(\rel X)[L]\to\graph(\rel A)$, for each $3$-element set $L$ containing $K$.
This finishes the proof that $\rel A$ does not have relational width $(2,l)$.
\end{proof}

We finally prove that for $k$ large enough, $\rel A$ has relational width $(k,k)$.
\begin{proposition}\label{prop:B-bounded-width}
	$\rel A$ has bounded relational width.
\end{proposition}
\begin{proof}
	Consider an instance $\rel X$ of $\csp(\rel A)$, and let $\mathcal P$ be a non-trivial compatible $(k,k)$-system for a large $k$.
	In particular, if $k\geq 3$ then every map $f\in P_K$, where $K\subseteq X$ with $|K|\leq k$, is a partial homomorphism $f\colon\graph(\rel X)[K]\to\graph(\rel A)$, as the unique relation of $\graph(\rel X)$ has arity $3$.
	
	Note that $\rel A$ is term-equivalent to the Boolean algebra $\rel B=(\{0,1\};\land,\lor,\neg,0,1)$, where the equivalence is given by the following identities:
	\begin{itemize}
	    \item $x\land y = (x\cdot x)\cdot (y\cdot y)$
	    \item $x\lor y = (x\cdot y)\cdot (x\cdot y)$
	    \item $\neg x = x\cdot x$
	    \item $0 = x\cdot (x\cdot x)$, $1=(x\cdot (x\cdot x))\cdot (x\cdot (x\cdot x))$.
	\end{itemize}
	By Lemmas~\ref{lem:eq} and~\ref{lem:general-reduction}, it is possible to compute an equivalent instance $\rel X'$ of $\csp(\rel B)$
	such that $\rel X'$ is itself a Boolean algebra.
	Let $k$ be large enough so that the identities defining a Boolean algebra, when translated into the signature of $\rel A$ using the identities above, do not have more than $k$ terms and subterms.\footnote{It can be seen that $k=9$ is large enough.}
	Then any two elements $a,b$ that are collapsed in the procedure described in Lemma~\ref{lem:eq}
	will be so that $P_{\{a,b\}}$ only contains constant maps:
	indeed, if $a$ and $b$ are collapsed because of the application of some identity $s\approx t$ (i.e., $a=s^\rel X(\mbf c)$ and $b=t^\rel X(\mbf c)$ for some $\mbf c$), one can let $K$ contain $a,b$ and all subterms of $s$ and $t$. Then $P_K$ only contains maps where $a$ and $b$ get the same value, since $P_K$ consists of partial homomorphisms, so that $P_{\{a,b\}}$ only contains constant maps.
	
	Finally, note that $\rel X'$ admits a homomorphism to $\rel B$ iff it has more than one element.
	Let $a\in X$ be arbitrary.
	Since $\mathcal P$ is non-trivial, $P_{\{a,a*a\}}$ is non-empty.
	Moreover, $P_{\{a,a*a\}}$ cannot contain any constant map since those are not homomorphisms $\graph(\rel X)[a,a*a]\to\graph(\rel A)$.
	Thus, it must be that $a$ and $a*a$ are not collapsed to a single element in $\rel X'$,
	and therefore $\rel X'$ has a homomorphism to $\rel B$.
	By Lemma~\ref{lem:general-reduction}, $\rel X$ has a homomorphism to $\rel A$, which concludes the proof.
\end{proof}

Note that $\graph(\rel A)$ does \emph{not} have bounded relational width, and in fact $\csp(\graph(\rel A))$ is NP-complete by Schaefer's theorem.
We mention that the structure $\rel Z:=(\{0,1\};x+y,x+y+1)$ with two binary operations is another example of a structure that has bounded relational width, that does not have relational width $(2,l)$ for any $l$, and whose graph $\graph(\rel Z)$ does not have bounded relational width.
The proof is similar as the one for $\rel A$.
Moreover $\csp(\graph(\rel Z))$ is the problem of solving linear equations over $\mathbb Z_2$, and is solvable in polynomial time.

We conclude this section with a few words about the notion of \emph{width},
which was introduced by Feder and Vardi in their seminal paper~\cite{Feder:1999}.
A template $\rel A$ is said to have width $(k,l)$ if the  $(k,l)$-consistency algorithm solves $\csp(\rel A)$ (as opposed to $(k,l)$-minimality solving $\csp(\rel A)$ for relational width).
We refer to~\cite{Feder:1999} for a description of the $(k,l)$-consistency algorithm.
For finite relational structures, it is known that every $\rel A$ that has bounded width has width $(2,l)$, where $l$ is taken to be at least 3 and as large as the arity of the relations of $\rel A$~\cite{BartoCollapse}.
The instance $\rel X$ and its compatible $(2,3)$-system $\mathcal P$ from the proof of Proposition~\ref{prop:not-23} also show that the template $\rel A=(\{0,1\};\cdot)$ does not have width $(2,3)$.
However, it can be seen that $\rel A$ has width $(2,l)$ for $l$ large enough, where the argument is the same as in Proposition~\ref{prop:B-bounded-width}.%

\section{Open problems}

Our results raise a number of important questions for the further development of a theory of CSPs of arbitrary  structures.

First, the known reduction from CSPs of relational structures to CSPs of algebras is a polynomial-time Turing reduction~\cite{Feder:2004}. Does a polynomial-time many-one reduction exist?

Secondly, we showed that for a finite structure $\rel B$, the complexity of $\csp(\rel B)$ only depends on the clone generated by the functions of $\rel B$, and on the partial polymorphisms of the relational reduct of $\rel B$, up to polynomial-time reductions. 
Is the complexity of $\csp(\rel B)$ captured by the polymorphisms of the relational reduct of $\rel B$?

Lemma~\ref{lem:eq} shows that for any \emph{finite} set of identities $\Sigma$ satisfied by a template $\rel A$ we can efficiently enforce $\Sigma$ in any instance $\rel X$ of $\csp(\rel A)$ without changing the number of homomorphisms. Is the same true for infinite sets of identities?

The algebras $\alg A$ for which we proved non-membership in $\Ksurj$ (Lemma~\ref{lem:bounded-arity} and Lemma~\ref{lem:exponentially-many-homos}) all have a counting function $\counting{\alg A}^s$ that grows as $2^{n^{\Omega(1)}}$.
In particular, we are not aware of any algebra $\alg A$ for which $\counting{\alg A}^s$ grows as, say, $n^{\log^k(n)}$.
Such an algebra would give rise to CSPs that can be solved in quasi-polynomial time, and their existence would be interesting from the point of view of complexity theory.
More generally, what growth regimes are possible for sequences of the form $\counting{\alg A}$ is a natural and interesting line of inquiry.

Finally, our results about the (relational) width of Boolean structures show a striking difference compared to finite relational structures.
Can one characterize bounded (relational) width algebraically?
If $\rel A$ has bounded width, does it have width $(2,l)$ for $l$ large enough?
Are there $k,l$ such that every structure $\rel A$ with bounded relational width has relational width $(k,l)$?

\bibliographystyle{plain}
\bibliography{refs}

\end{document}